# 3D-Printing for Analytical Ultracentrifugation (preprint)

Abhiksha Desai[1], Jonathan Krynitsky[2], Thomas J. Pohida[2], Huaying Zhao[1] and Peter Schuck[1]*

[1]Dynamics of Macromolecular Assembly Section, Laboratory of Cellular Imaging and Macromolecular Biophysics, National Institute of Biomedical Imaging and Bioengineering, National Institutes of Health, Bethesda, Maryland 20892

[2]Division of Computational Bioscience, Center for Information Technology, National Institutes of Health, Bethesda, Maryland 20892

*for correspondence and proofs:  schuckp@mail.nih.gov

## Abstract

Analytical ultracentrifugation (AUC) is a classical technique of physical biochemistry providing information on size, shape, and interactions of macromolecules from the analysis of their migration in centrifugal fields while free in solution. A key mechanical element in AUC is the centerpiece, a component of the sample cell assembly that is mounted between the optical windows to allow imaging and to seal the sample solution column against high vacuum while exposed to gravitational forces in excess of 300,000 g. For sedimentation velocity it needs to be precisely sector-shaped to allow unimpeded radial macromolecular migration. During the history of AUC a great variety of centerpiece designs have been developed for different types of experiments. Here, we report that centerpieces can now be readily fabricated by 3D printing at low cost, from a variety of materials, and with customized designs. The new centerpieces can exhibit sufficient mechanical stability to withstand the gravitational forces at the highest rotor speeds and be sufficiently precise for sedimentation equilibrium and sedimentation velocity experiments. Sedimentation velocity experiments with bovine serum albumin as a reference molecule in 3D printed centerpieces with standard double-sector design result in sedimentation boundaries virtually indistinguishable from those in commercial double-sector epoxy centerpieces, with sedimentation coefficients well within the range of published values. The statistical error of the measurement is slightly above that obtained with commercial epoxy, but still below 1%. Facilitated by modern open-source design and fabrication paradigms, we believe 3D printed centerpieces and AUC accessories can spawn a variety of improvements in AUC experimental design, efficiency and resource allocation.

## Introduction[1]

Analytical ultracentrifugation (AUC) is a classical technique of physical biochemistry for the study of size, shape, and interactions of macromolecules free in solution through the application of a strong gravitational force and the real-time observation on resulting redistribution of macromolecular concentration [1–4]. In the last decades, it has undergone significant development with modern instrumentation, theoretical approaches, and computational analysis methods [5–13]. It has wide-spread applications in a range of fields including structural biology [14,15], macromolecular hydrodynamics [16–18], supra-molecular chemistry [19], food science [20], biomaterials [21], nanoparticles [22–24] and protein pharmaceuticals [25–28]. In the study of reversible protein complex formation, AUC offers unique opportunities to measure the energetics of multi-protein complexes with dissociation equilibrium constants ranging from picomolar to millimolar [29–31], and exploiting multi-wavelength analysis to determine binding stoichiometry of multi-protein complexes [9,14,32–35].

A key element in AUC equipment is the centerpiece, a sample holder consisting of an arrangement of two sector-shaped cuvettes embedded in an epoxy resin or metal cylinder that is sandwiched between quartz or sapphire windows, and mounted into a cylindrical barrel [36]. This assembly is inserted into the rotor, allowing real-time optical detection of radial macromolecular migration in the spinning rotor during the AUC experiment. The centerpiece assembly needs to seal the sample solution against evaporation in the high vacuum of the rotor chamber, and provide mechanical stability at up to ~300,000 g. Also, it needs to permit sufficient heat conduction for thermal equilibration in order to avoid thermal convection, and, for sedimentation velocity experiments, it needs to be sector-shaped so as to allow un-impeded radial migration of macromolecules. While the basic centerpiece concept was developed by Svedberg in the 1920s [37] and improved by Pickels in the 1940s [36], a large number of variations have been developed over the decades. For example, centerpieces have be designed to provide different optical path lengths and/or sample volumes, special solution column geometries, additional sample compartments, trap-doors or capillaries for liquid flow at the start of the AUC experiment, additional mechanical elements facilitating fractionation, altered chemical resistance and improved thermal properties through use of different materials, and features enabling different filling and cleaning techniques [38–45].

---

[1] Abbreviations: AUC: analytical ultracentrifugation; SE: sedimentation equilibrium; SV: sedimentation velocity; EGFP: enhanced green fluorescent protein; BSA bovine serum albumin

Despite many reasons for variations in centerpiece design such that they facilitate particular AUC experiments, a significant hurdle in the implementation of new centerpieces designs is their fabrication. This is largely because the required prototyping or manufacturing capability with sufficient precision is not easily accessible to most laboratories, and the process usually requires iterative improvement and is expensive. Even the purchase of commercially available standard centerpieces is associated with high cost in excess of $1,000. Finally, a problem is their limited life-time due to scratches and deformation in the centrifugal field with time, which is why they are regarded as consumables.

In recent years the power and accessibility of 3D printing has dramatically increased, and novel applications have arisen in many fields, including open-source optics and other laboratory equipment [46–49]. Several developments in 3D printing are particularly promising when considering potential application in the fabrication of AUC centerpieces: (1) A large variety of materials are now available, including very strong polymers and several metals, such as titanium, previously used for centerpieces [44]; (2) The dimensional accuracy and resolution of 3D printed objects is steadily increasing, with layer thickness and lateral resolution in the low micrometer range; (3) homogeneous non-permeable parts with good surface quality are more easily achieved; and (4) 3D printing is low-cost and has become readily accessible, even without investment in obtaining in-house printers, through many web-based mail-order printing services.

For these reasons, we have explored the application of 3D printing technology for the manufacturing of AUC system components. We demonstrated that centerpieces suitable in terms of mechanical stability can be fabricated with 3D printing at a small fraction of the cost of purchasing off-the-shelf commercial centerpieces. Carrying out AUC experiments with standard reference molecules, we found these centerpieces to be sufficiently precise to permit both sedimentation equilibrium (SE) experiments (the observation of thermodynamic equilibrium of macromolecular distribution at typically 5,000 – 50,000 g), and sedimentation velocity (SV) experiments (the analysis of the dynamics of the sedimentation process typically at 200,000 – 300,000 g). Furthermore, ancillary AUC components may be 3D printed, such as cell assembly holders for iButton probes that measure the temperature of the spinning rotor [50], window holders, and spacer rings. Finally, 3D printing allows the straightforward creation of novel centerpieces with, for example, custom optical path lengths and sample volumes to improve the dynamic range of the optical detection. In some cases, centerpiece fabrication with 3D printing enables innovative designs and configurations that could not have been considered before due to limitations of traditional manufacturing methods. In summary, we believe 3D printing of AUC components will allow more efficient use of laboratory resources, both in funds and in sample material, and opens possibilities for innovative AUC experiments.

## Materials and Methods

### Model Design and 3D Printing

All objects were 3D designed using OpenSCAD, which is freely available at http://www.openscad.org/. Our custom AUC centerpiece OpenSCAD model files will be available in the resources section of our laboratory website https://sedfitsedphat.nibib.nih.gov/tools/default.aspx, and may be used as templates for further design modification. For 3D printing, the model files were exported to the stereolithography (STL) files, and will be available at the NIH 3D Print Exchange (http://3dprint.nih.gov/). The STL files were submitted to different web-based 3D printing services, as well as printed in-house, using different printer technologies, materials, and manufacturers.

Centerpieces in acrylic (VeroClear-RGD810) were 3D printed in our laboratory on an Eden260vs PolyJet (Stratasys, Eden Prairie, MN). Printing took 50 – 60 min, dispensing 16 μm layers onto a build plate, immediately cured by a UV lamp. Water-soluble support material was removed initially with a water-jet station (a power washer in a contained chamber), followed by a bath in a 2% sodium hydroxide solution for at least 5 hours, and completed with a final water-jet wash to remove any remaining support material residue. In principle, the soluble support material offers the ability to directly create curved fluid channels and moveable components.

Carboxylate centerpieces were 3D printed by stereolithography from liquid photopolymer cured in layers by laser exposure (i.e., SLA 3D printing). Centerpieces in "prime gray" were printed at i.materialise.com. Prime gray is a proprietary material, similar in appearance and mechanical strength to 3,4- Epoxycyclohexylmethyl 3,4-epoxycyclohexane carboxylate (Accura Xtreme, 3D Systems Inc., Rock Hill, SC.). Polypropylene carbonate centerpieces "Accura Xtreme White 200" were printed in 0.004'' layers at buildparts.com (C.Ideas, Crystal Lake, IL), and in 0.002'' layers at protolabs.com/fineline (Proto Labs Inc., Maple Plain, MN). Centerpieces from a proprietary ABS-like resin, "MicroFine Green", were printed by micro-stereolithography in 0.001'' layers also at protolabs.com/fineline. The post-print processing steps, such as support material removal, are generally unknown when utilizing web-based 3D printing services; in the present case only 'standard' treatment was ordered.

**Analytical Ultracentrifugation**

AUC experiments were carried out in an Optima XL-A (Beckman Coulter, Indianapolis, IN), calibrated as previously described [50,51]. Data were acquired using the installed absorbance detection, or, alternatively, a confocal fluorescence detection system (FDS, Aviv Biomedical, Lakewood NJ) equipped with either a 10 mW solid state laser exciting at 488 nm and emission bandpass filter from 505 nm to 565 nm, or an adjustable diode laser exciting at 561 nm and a long-pass emission filter at 580 nm. Unless otherwise mentioned, setup of the AUC followed standard protocols [52], using an 8-hole An-50 TI rotor. The 3D printed centerpieces were sealed with a standard gasket (part # 330446, Beckman Coulter, Indianapolis, IN) against sapphire or quartz windows, unless otherwise mentioned.

For the sedimentation velocity (SV) experiments, 400 µL of a solution of bovine serum albumin (BSA) at 1.1 mg/ml dissolved in phosphate buffered saline (PBS) was placed in the sample sector, and 400 µL of PBS in the reference sector, forming the respective solution columns after rotor acceleration. After temperature equilibration at 20 °C, the rotor was accelerated to 50,000 rpm, and data were acquired at 280 nm. Data were analyzed in SEDFIT using the $c(s)$ model [53]. For the sedimentation equilibrium experiment, 150 µL enhanced green fluorescent protein (EGFP) at 7.7 µM in PBS and an equal volume PBS was filled in the sample and reference sectors, respectively. A time-optimized rotor speed protocol [13] was used to attain equilibrium sequentially at 15,000 rpm, 24,000 rpm, and 35,000 rpm, acquiring absorbance data at 489 nm. Data analysis was carried out in SEDPHAT using a single non-interacting species model allowing for TI and RI noise [54].

# Results

We first evaluated model materials, including different plastics and metals, for suitability in 3D printing of AUC centerpieces. Based on the smoothest surfaces impermeable to water, we initially selected carboxylate material "prime gray" for further testing of the centerpiece mechanical stability in the AUC. In order to minimize the stresses the centerpieces must sustain in the centrifugal field, we chose the outer diameter of the centerpiece such that a tight fit in the cylindrical aluminum barrel of the cell assembly was achieved. Approximately sectorial holes were designed to hold liquid samples in the standard position for transmission of light during rotation in the rotor. A seal against the optical windows was readily achieved with polyethylene gaskets (see below). With water-filled sectors, the assembly was placed in the rotor and exposed to increasingly higher centrifugal fields in several steps testing for damage to the centerpiece after each step. Unexpectedly, no breakage, deformation, or sample leak could be discerned even at the highest rotor speed of 60,000 rpm, which corresponds to a gravitational field of > 300,000 g at the highest radius.

Encouraged by this result, we printed centerpieces in different materials, and examined their mechanical stability at rotor speeds of up to 50,000 rpm, which is the maximum rotor speed for 8-hole rotors and applied in our standard protocol [2,52]. For acrylic centerpieces, the centerpiece orientation during 3D printing had a large effect on mechanical durability. When the solution column inner radial walls were positioned approximately parallel to the build plate, intended to make smother column walls by taking advantage of higher accuracy of the printer in z-direction than x- and y-direction, the centerpieces broke in the solution column corner at the smaller radius in the centrifugal field. However when centerpiece was oriented so the printing build plate was parallel to the plane of rotation, no breakage occurred. Thus, all subsequent printing of centerpieces was done in this orientation. Another mechanical failure was encountered with an experimental acrylic centerpiece design featuring parallel sample walls creating a rectangular solution column (see below). After use for several days at 50,000 rpm, the rectangular solution column centerpiece exhibited strong permanent deformations, though no breakage. This was not observed with acrylics centerpieces with the standard sector-shaped design in a side-by-side control experiment in the same run. Some polycarbonate centerpieces were found to deform after several runs at 50,000 rpm totaling 30 – 40 hours, leaving the central divider bent.

In the process of acquiring different centerpieces for testing we obtained centerpieces printed with different layer thicknesses. For example, polycarbonate "Accura Xtreme White" centerpieces were printed in either 100 μm or 50 μm thickness, while a proprietary resin "MicroFine Green", was printed in 25 μm layers (**Figure 1**). In our observation this had no discernable impact on the performance of the centerpiece in AUC experiments described below.

The centerpieces require a vacuum seal against the sapphire or quartz window of the cell assembly to prevent evaporation of the liquid samples in the high vacuum of the rotor chamber. In commercial epoxy resin centerpieces the seal is achieved through sufficient flatness of the surfaces and centerpiece compliance, whereas for the commercial aluminum and titanium centerpieces polyethylene or Teflon gaskets are used [44]. Such gaskets were required for all 3D printed centerpieces except for "MicroFine Green" centerpieces, which exhibited top and bottom surfaces in the majority of cases sufficiently smooth to seal against an optical sapphire window after torqueing the cell assembly barrel to the standard 120 – 140 inch/lbs. The seal was observed to degrade in subsequent runs, however, requiring gaskets. An improved seal was achieved with a design with an embossed ridge of 100 μm height and 300 μm width on the surface adjacent the sectors. Surfaces with such an embossed gasket were also found to create a seal when employed with polycarbonate centerpieces.

After having established the mechanical stability and vacuum seal of the 3D printed centerpieces, we next explored the possibility of using the new centerpiece for sedimentation equilibrium (SE) experiments. SE experiments are the least mechanically challenging type of AUC experiments due to the low rotor speeds, and due to the independence of the shape of the Boltzmann distribution in equilibrium (or its exponent, corresponding to the molecular weight) on solution column geometry [2]. **Figure 2** shows SE data of an EGFP sample acquired with absorbance optics at a sequence of different rotor speeds, which can be modeled very well with the expected Boltzmann distributions corresponding to an apparent molar mass of 29.7 (27.4 – 32.1) kDa. For comparison, a standard Epon double sector cell with the identical sample in the same run led to a best-fit apparent molar mass of 27.3 (25.2 – 29.5) kDa.

Sedimentation velocity (SV) experiments pose more stringent demands on the centerpiece geometry to permit convection-free sedimentation and are typically carried out at higher centrifugal fields. Encouraged by the results of the centerpiece stability tests and the SE pilot experiments, a second generation centerpiece was designed to be precisely sector-shaped. As in standard calibration experiments of a recent multi-laboratory study [51], we carried out an SV experiment with BSA as a reference sample, and analyzed the absorbance data from the sedimentation boundaries arising from molecular sedimentation at 50,000 rpm with the sedimentation coefficient distribution $c(s)$. The familiar hydrodynamic resolution of different BSA oligomers was observed. With the "prime gray" centerpiece, a monomer s-value of 4.24 S was obtained with an apparent monomer molar mass of 58.1 kDa, whereas a standard Epon centerpiece control in the same run led to 4.28 S and 59.2 kDa (**Figure 3**). Both values are well within the values obtained in a multi-laboratory benchmark study of (4.30 ± 0.19) S [51]. This demonstrates that AUC SV experiments are possible with 3D printed centerpieces with reasonable precision.

Centerpieces 3D printed with different technologies and materials show virtually identical sedimentation boundaries and very similar peaks in the sedimentation coefficient distributions (**Figure 3B**). The consistency is remarkable considering that the detailed peak heights and widths in c(s) distributions are determined jointly by the data signal/noise ratio and mathematical analysis (scaling of regularization), which are usually not well reproducible from experiment to experiment independent of centerpieces [52]. In a replicate experiment with four identical polycarbonate centerpieces containing the same BSA solution side-by-side in the rotor, a statistical precision of the BSA monomer *s*-value of 0.96% was obtained. This is higher than the standard deviation of 0.15% previously observed for BSA monomer in Epon centerpieces when studied side-by-side in the same run [50]. The reproducibility of the apparent molar mass was 1.9%.

With the goal to assess the sensitivity of the SV experiment to the precise centerpiece geometry, as a negative control we 3D printed an acrylic centerpiece with parallel walls, generating a rectangular solution column. Molecules migrate radially in the centrifugal field, thus exhibiting a velocity component perpendicular to the non-radial walls in the parallel-walled centerpiece. The non-radial velocity component is expected to create lateral density gradients that lead to convection [2–4,36,37,55]. We carried out a BSA sedimentation experiment at 50,000 rpm with rectangular and sectorial centerpieces from the same material side-by-side in the rotor. It can be discerned from **Figure 4** that the rectangular solution still allows sedimentation boundaries to form, although with a disturbance of the boundary height and the absence of a plateau. These results suggest the effects of a rectangular cell geometry counteracts the radial dilution usually associated with macromolecular migration in the centrifugal field. The rectangular cell geometry causes a significantly lower quality of fit with the standard $c(s)$ sedimentation model, and the best-fit boundary shows higher signals than in the sectorial cell geometry. Remarkably, however, the resulting sedimentation coefficient distributions virtually superimpose each other (**Figure 5**), still exhibiting baseline resolved monomer and dimer at the correct sedimentation velocities. Notably, the rectangular cell geometry generated increased boundary broadening, which, when interpreted as diffusion in the $c(s)$ analysis, resulted in a 10.2% lower estimates of the apparent molar mass of the monomer. Thus, while sector-shaped solution columns are important for precise AUC SV experiments as expected, these data demonstrated there is also a surprising tolerance of the major boundary features to imperfections in the solution column.

Finally, as an example for the ease of centerpiece design afforded by 3D printing technology, we created a centerpiece for use specifically in conjunction with the confocal fluorescence detection system (FDS). Since this detection requires optical access only through a single upper window, the centerpiece was designed with increased height replacing both the lower window and the customary spacers otherwise required for 3 mm centerpieces. Sector-shaped wells of 3 mm depth were created at the top of the centerpiece and connected with diagonal filling holes to the fill ports of the standard aluminum barrel. To facilitate loading without air locks separate venting channels were included, and an embossed gasket was added to the top surface for the 3D printed centerpiece to self-seal (inset in **Figure 6B**). Similarly, a FDS calibration centerpiece was created, also featuring filling and venting holes, as well as an embossed seal. Fluorescence data of mCherry sedimenting at 50,000 rpm acquired with this centerpiece are shown in **Figure 6**. They exhibit the characteristic signal magnification gradient of FDS data acquired at shallow focal depths which can be computationally accounted for [56]. The main species sediments at 2.68 S with a molar mass of 26.9 kDa which compares well to the value of 28.9 kDa expected from amino acid sequence [57].

## Discussion

The present communication describes the novel concept of fabricating AUC centerpieces by 3D printing. We were surprised to find that these centerpieces can be sufficiently mechanically stable to withstand prolonged exposure to gravitational fields in excess of 300,000 g, while sealing the solution against high vacuum of the rotor chamber. Furthermore, we found them to be precise enough to allow a variety of AUC experiments. We believe this has several practical implications.

First, AUC centerpiece 3D print fabrication offers a considerable reduction in cost. Centerpieces can be made by various printing technologies (e.g., Polyjet, SLA) and numerous materials, including acrylic, polycarbonate, carboxylate, nylon, and several metals, all between < 1% to 10 % of the cost of commercial centerpieces. At present we have very little experience with the long-term durability of the non-metallic centerpieces, which appears to be limited, for example, in the polycarbonate centerpieces. However, this may not be a critical factor considering the low cost, and may not be significant for low-field applications in SE. Ironically, with many of the centerpiece materials we tested, the gasket used to provide a seal between the centerpiece and optical window is now the higher cost consumable component. However, we found the need for a gasket may be eliminated by improved centerpiece design, using higher resolution 3D printing (i.e., better surface quality), and careful material selection.

The selection of material may be guided by the experimental requirements for chemical compatibility. With some 3D printing technologies and corresponding materials, unreacted polymer, solvents, or plasticizers may leach into the AUC sample solution within the experiment timeframe. The interference optical detection system offers a convenient means for detection of 3D printing impurities in the sample sector. In one instance, we have observed such effects with a freshly 3D printed carboxylate centerpiece over the course of several days of incubation (a time-frame often required for SE experiments). Related, surface adsorption can be a problem when studying protein interactions in the sub-nanomolar range with fluorescence detection [30], and further studies will be required to gain experience with beneficial or detrimental properties of different materials in this and other applications of AUC.

3D printed centerpieces can support a wide range of possible AUC applications. Since sedimentation equilibrium only requires mechanical stability of the solution column over long time, the new centerpieces should be straightforward to apply without further consideration. Fabrication of new centerpieces with different optical path lengths, along with suitably sized spacer rings, is a routine and inexpensive process. The independence of equilibrium from the solution column shape can be exploited in the design and fabrication of non-sectorial centerpieces to shorten the time to equilibrium [3]. Multi-welled centerpieces accommodating multiple samples per cell, similar to commercially available 6-channel centerpieces, are similarly conceivable.

Perhaps the most surprising result of the present work was the successful use of 3D printed centerpieces in high-quality SV AUC experiments; with results for the BSA monomer virtually identical to those carried out in commercial epoxy centerpieces and within the limits of the large multi-laboratory benchmark study recently published [51]. We did observe significantly higher standard deviation of the BSA monomer sedimentation coefficient, but variation in replicate experiments was still less than 1%. This suggests that 3D printed centerpieces examined in the present work may not yet be suitable for SV experiments requiring the highest possible precision, but this leaves a remarkable range of useful and innovative SV applications.

A pilot experiment with a rectangular centerpiece – grossly violating known requirements for unimpeded sedimentation and designed to produce strong convection – was astonishingly successful in producing the correct sedimentation coefficient distribution. This experiment suggests that the shape of the sedimentation boundary, specifically the degree of boundary broadening, is the feature in SV most sensitive to unwanted convection, rather than the average migration velocity of the boundary. We hypothesize that over-concentration of protein at the parallel side-walls is effectively opposed by lateral diffusional flows, and that this also alleviates the effects of minor imperfections in the sector wall smoothness of any 3D printed centerpieces.

The determination of whether 3D printed centerpieces using current material and technology are of sufficient quality to carry out SV experiments may be governed by the question posed in the experiment. For example, trace aggregate detection of small and medium sized proteins puts higher demands on precision than the determination of $s$-values of the main components [58], whereas protein interaction analysis *via* isotherms of the boundary structure and $s$-values in the $s_w$ analysis [59,60] and in the effective particle model [61,62] disregards features of boundary shape entirely. Similarly, 3D printed centerpieces may be most suitable for AUC teaching environments thus facilitating the dissemination of the technology.

The ease of centerpiece 3D print fabrication promotes experiment creativity, and ultimately the development of novel AUC methods. For example, historically, a variety of centerpieces with additional sample reservoirs and channels for transfer of liquid after start of centrifugation have been conceived, and the 3D print fabrication of well-defined channels may allow more complex layering techniques. Also, in view of our experiment with the parallel-walled solution column, it is worth considering that much of our understanding of the influence of imperfections in centerpieces on macromolecular sedimentation behavior has been derived from case studies with centerpieces exhibiting poorly defined features, such as accidental file marks or scratches observed to cause aberrant boundaries [36,55], or from unspecified improvements in manufacturing [63]. Here, 3D printing offers the new opportunity to fabricate centerpieces cheaply and reproducibly with features of well-defined geometries to enable more systematic studies of the conditions necessary for convection-free sedimentation in SV. Possible improvements in the precision of SV measurements could stimulate advances in our understanding of molecular hydrodynamics and protein solvation [64,65], and are of great practical interest in the characterization of protein pharmaceuticals in biotechnology [63]. Finally, the utility of 3D printing in AUC is not limited to centerpiece fabrication. For example, it is possible to print a steel or titanium rotor hole inserts that accommodate iButton temperature loggers to permit temperature measurements of the spinning rotor [50], whereas, without access to a machine shop for custom fabrication of such a holder, temperature monitoring was previously restricted to measurements in the resting rotor [66].

In conclusion, we believe the rapid and low-cost prototyping and open-source design of novel and functional AUC centerpieces and other system components afforded by 3D printing technology opens significant new opportunities by both enabling the development of new AUC methodologies for specialized and emerging applications, and improving efficiency and ultimately precision of existing AUC configurations.

## Acknowledgment

This work was supported by the Intramural Research Program of the National Institute of Biomedical Imaging and Bioengineering, National Institutes of Health, United States. We thank Dr. George Patterson for the samples of EGFP and mCherry.

## Figure Legends

**Figure 1:** Picture of a 12 mm pathlength centerpiece printed of ABS-like resin MicroFine Green. Sedimentation velocity data of BSA at 50,000 rpm collected with this centerpiece installed into a standard cell assembly without gaskets are shown in Figure 3B.

**Figure 2:** Radial concentration distribution in a sedimentation equilibrium experiment with enhanced green fluorescent protein in a "prime gray" photopolymer centerpiece. Data were acquired with the absorbance detection sequentially at rotor speeds of 15,000 rpm (purple), 24,000 rpm (blue), and 35,000 rpm (cyan) (symbols, only every $5^{th}$ data point shown). A global model (lines) results in an apparent molar mass of 29.7 kDa with a root-mean-square deviation (rmsd) of 0.0032 $OD_{489}$, and residuals as shown in the lower plot.

**Figure 3:** Temporal evolution of radial concentration profiles in a sedimentation velocity experiment with bovine serum albumin in a "prime gray" photopolymer centerpiece. *Panel A:* Absorbance data acquired at a rotor speeds of 50,000 rpm at a series of time points (symbols, only every $3^{rd}$ data point of every $2^{nd}$ scan shown, with color temperature indicating progress of time). The $c(s)$ fit (lines) results in an rmsd of 0.0065 $OD_{280}$, with the residuals shown in the small plots as residuals bitmap and superposition. *Panel B:* The corresponding $c(s)$ distribution (magenta), and for comparison the $c(s)$ distribution from a control in the same run using a standard Epon centerpiece (black); microgreen (green); Xtreme white (blue dashed); in-house clear (cyan dotted).

**Figure 4:** Sedimentation velocity analysis of bovine serum albumin sedimenting at 50,000 rpm in acrylic centerpieces with a sector-shaped (A) and rectangular shaped (B) solution column. The protein sample was identical in both. The upper panel shows the sedimentation boundaries (points, for clarity, only every $2^{nd}$ data point of every $2^{nd}$ scan is shown), along with the best-fit c(s) profiles (solid lines). Below are the residuals of the fit as bitmap and overlay plot. The $c(s)$ distribution for both data sets are shown in Figure 5.

**Figure 5:** Sedimentation coefficient distributions calculated from the data in Figure 4 for rectangular (magenta) and sectorial (blue) geometry.

**Figure 6:** Fluorescence optical data in a 3D printed carbonate centerpiece featuring a 3 mm deep sector-shaped well at the top, filling and venting holes, and an embossed seal. The focal depth of the fluorescence optics was 2.0 mm. (A) Shown are sedimentation profiles acquired with 561 nm excitation for 46 nM mCherry [57] dissolved in phosphate buffered saline (dots), and best-fit $c(s)$

sedimentation coefficient distribution with adjustments for characteristic signals of fluorescence detection [56] (solid lines). The plot appended below shows the residuals of the fit. (B) Corresponding sedimentation coefficient distribution showing a main peak at 2.68 S and diffusional boundary broadening corresponding to a species of 26.9 kDa.

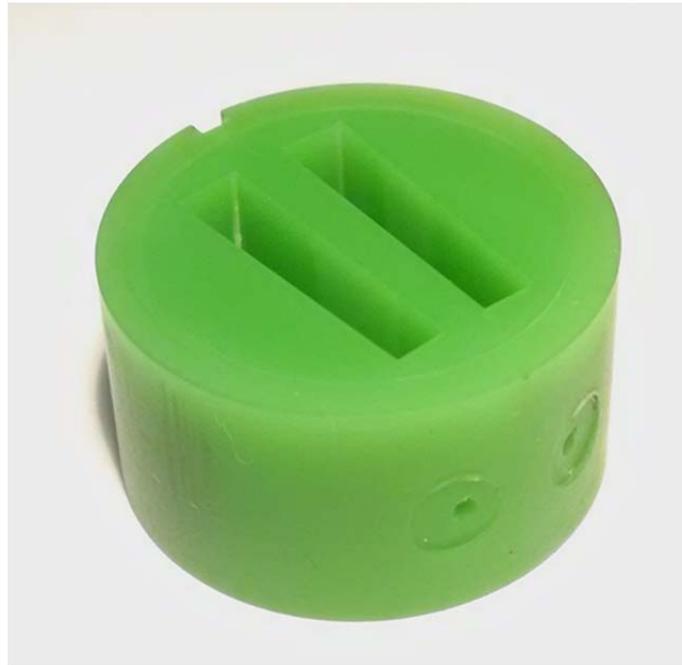

**Figure 1**

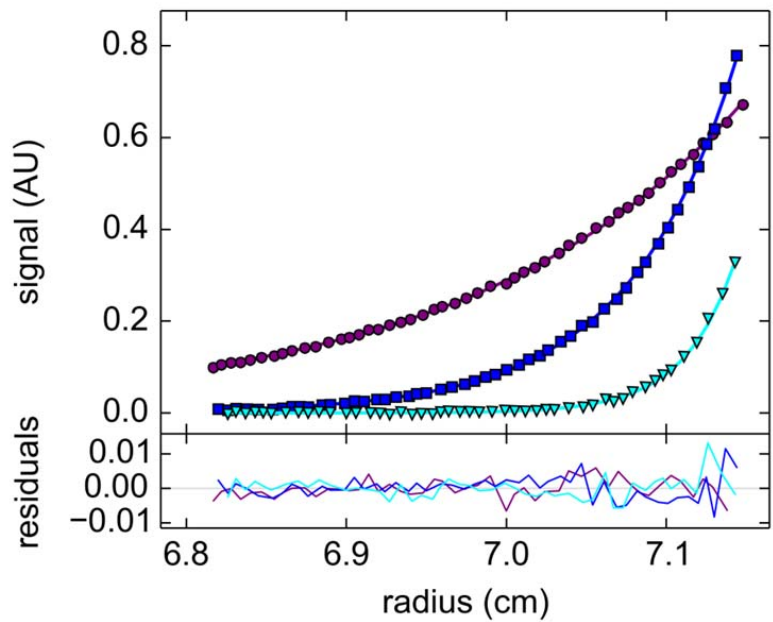

**Figure 2**

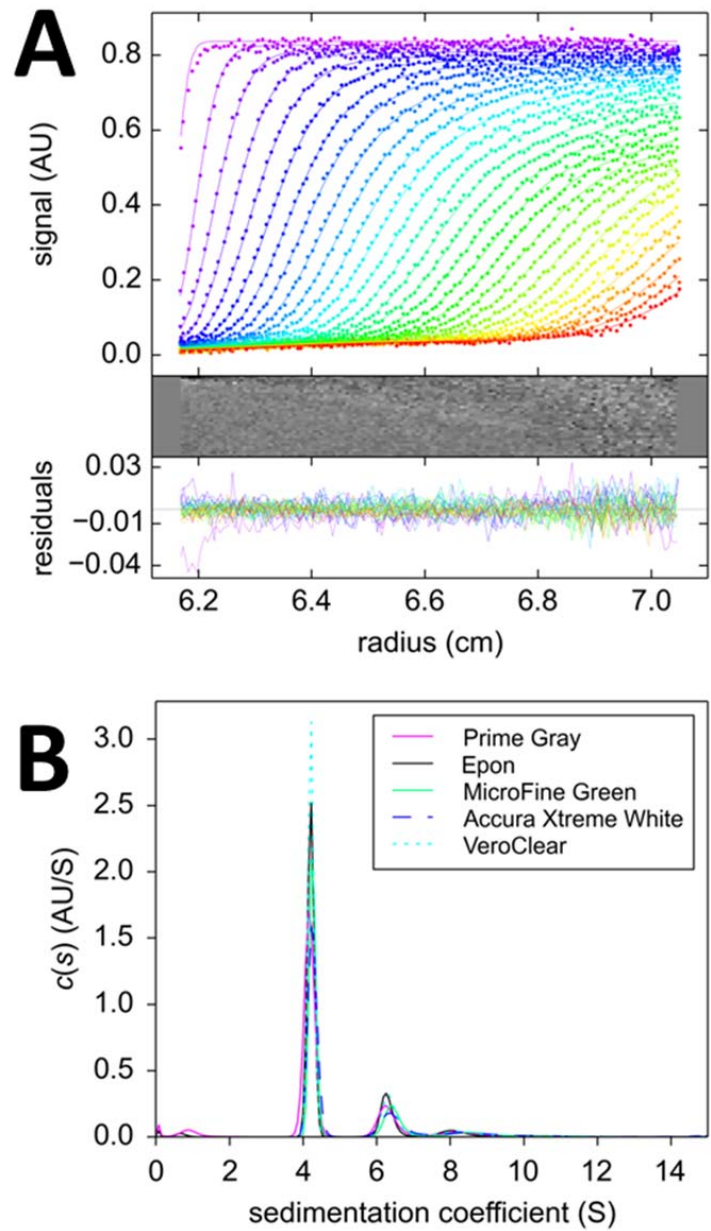

**Figure 3**

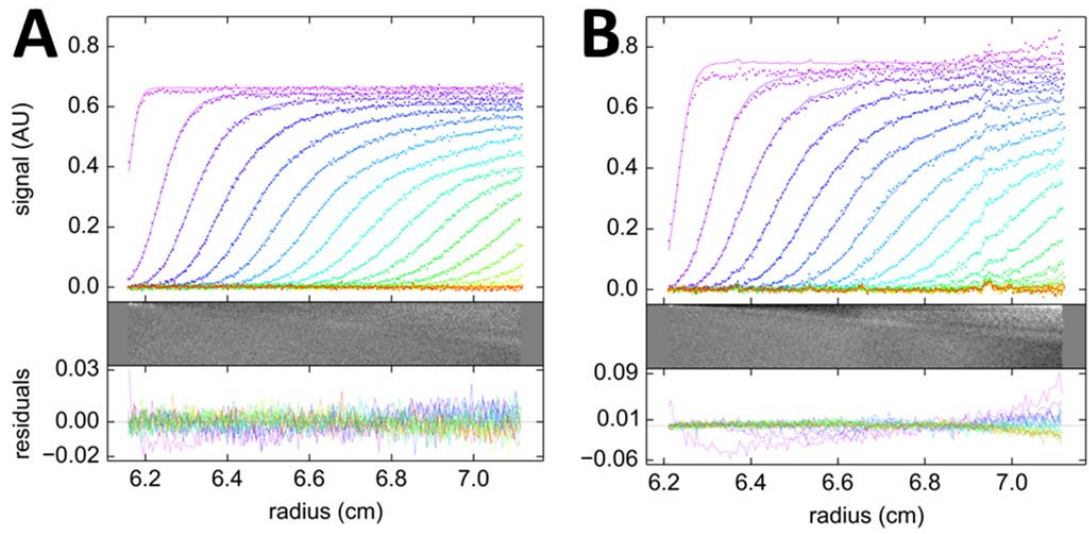

**Figure 4**

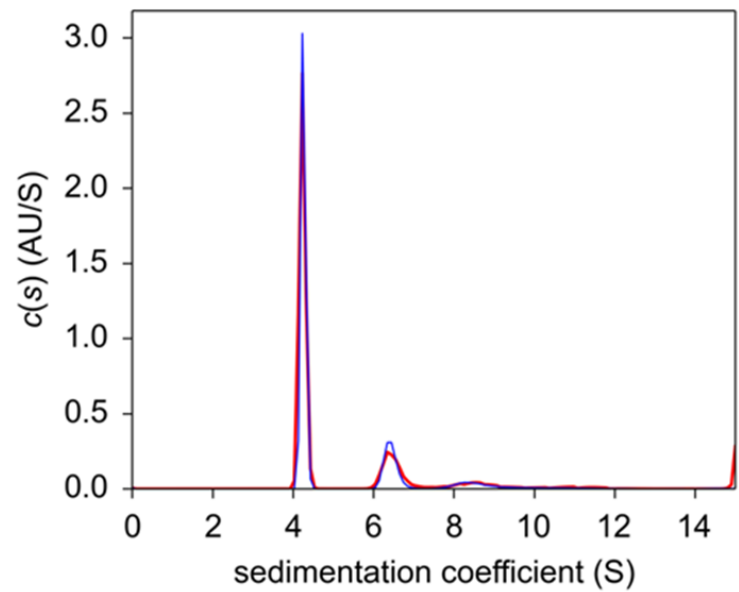

**Figure 5**

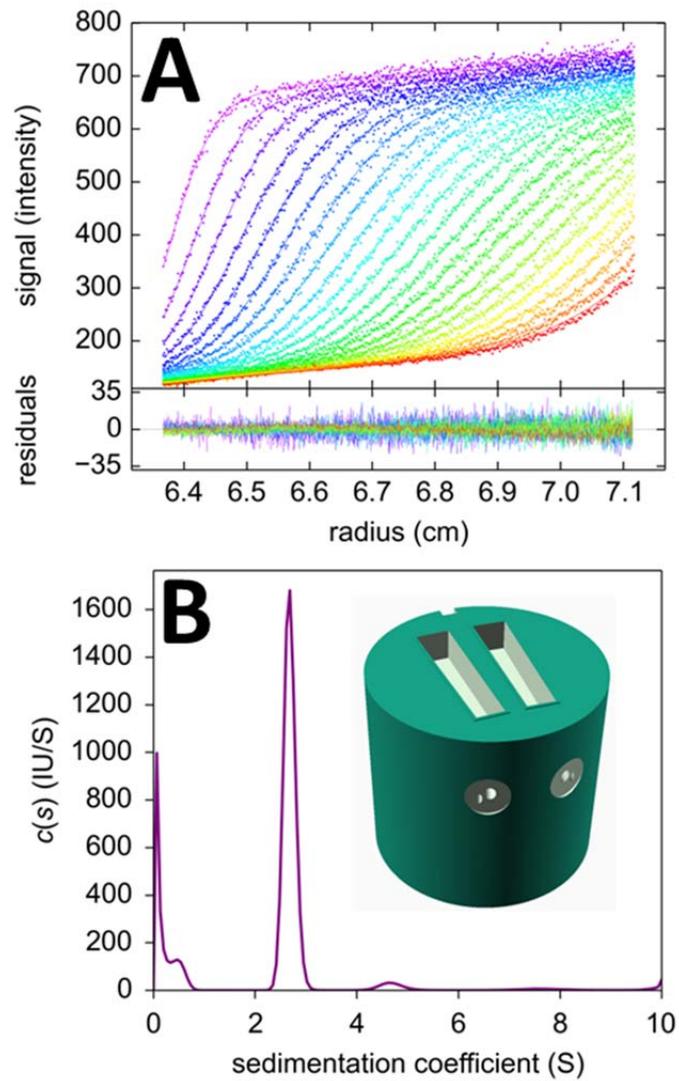

**Figure 6**